\documentclass[%
 reprint,
 amsmath,amssymb,
 aps,
]{revtex4-2}
\usepackage[colorlinks,linkcolor=blue,anchorcolor=blue,citecolor=blue,filecolor=blue,menucolor=blue,runcolor=blue,urlcolor=blue,frenchlinks=blue]{hyperref}
\usepackage{amsmath}
\usepackage{graphicx}
\usepackage{amssymb}
\usepackage{dcolumn}
\usepackage{mathrsfs}
\usepackage{bm}
\usepackage{braket}
\usepackage[T1]{fontenc}
\usepackage{color}
\definecolor{red}{RGB}{255,0,0}

\begin{document}

\title{Photo-ionization Compensation of Stray Electric Fields for Cold Rydberg Atoms}

\author{Zi-Yuan Chen}
\thanks{\hypertarget{eqcontrib}{These authors contributed equally to this work.}}

\author{Zhao-Xin Fu}
\thanks{\hypertarget{eqcontrib}{These authors contributed equally to this work.}}

\author{Ze-Rui He}
\thanks{\hypertarget{eqcontrib}{These authors contributed equally to this work.}}

\author{Zhi-Yong Chen}
\thanks{\hypertarget{eqcontrib}{These authors contributed equally to this work.}}

\author{Si-Ao Cheng}

\author{Jia-Hao Liang}
\altaffiliation{Present address: MatriQ Co.,Ltd.}

\author{Shan-Chao Zhang}

\author{Yan-Xiong Du}
\email{yanxiongdu@m.scnu.edu.cn}

\author{Chang Li}
\email{lichangphy@gmail.com}

\affiliation {Key Laboratory of Atomic and Subatomic Structure and Quantum Control (Ministry of Education), School of Physics, South China Normal University, Guangzhou 510006, China}

\affiliation {Guangdong Provincial Key Laboratory of Quantum Engineering and Quantum Materials, South China Normal University, Guangzhou 510006, China}

\affiliation {Guangdong-Hong Kong Joint Laboratory of Quantum Matter, Frontier Research Institute for Physics, South China Normal University, Guangzhou 510006, China}


\begin{abstract}
Neurtal atoms in optical tweezer arrays constitute a highly promising platform for quantum computing and quantum simulation. Their operation relies on precise control of the Rydberg excitation, which is highly sensitive to background electric fields. Here, we identify a previously overlooked source of stray electric fields arising from trapped charges within the antireflection coating layers of glass vacuum cells. In contrast to the conventional approach of removing surface charges through ultraviolet-light-induced desorption, we compensate these clamped charges by generating additional charges via photo-ionization of a cold atomic ensemble. We verify the resulting suppression of stray electric fields through Rydberg excitation spectroscopy in an atomic array and further confirm that the residual electric field inside the vacuum cell is effectively eradicated with the aid of external electrodes. Our work identifies and mitigates a previously unrecognized source of residual electric field, providing a practical solution for improving the performance of neutral atomic quantum processors and other Rydberg-based quantum technologies, as well as surface-sensitive atomic systems.

\end{abstract}

 \maketitle

\section{introduction}
\label{1}

Neutral-atom arrays have emerged as a versatile and highly promising platform for quantum computing and quantum simulation, where strong Rydberg-mediated interactions between individually trapped atoms provide the essential resource for quantum control. This platform has enabled a series of remarkable advances, including high-fidelity entangling gates and logical quantum processors~\cite{PhysRevLett.123.170503, PhysRevA.105.042430, graham2022multi, 2023High, xu2024constant, bluvstein2024logical, evered2026highfidelityentanglinggatesnonlocal}, as well as programmable quantum simulation of many-body quantum systems~\cite{RevModPhys.82.2313, 2015Entangling, 2017Probing, doi:10.1126/science.abi8794, 2022Continuous, 2024Observation, w1cp-l5vq}. Central to these advances is the ability to coherently excite atoms to Rydberg states, which determines the fidelity and scalability of subsequent quantum operations.
However, owing to their large principal quantum numbers ($n \gg 1$), Rydberg atoms possess an enormous electric polarizability that scales as $n^7$~\cite{gallagher1994rydberg}, rendering them exceptionally sensitive to ambient electric fields. Even weak stray electric fields can induce significant Stark shifts of the Rydberg transition, spectral broadening, and rapid dephasing, thereby degrading the coherence of ground–Rydberg transitions and ultimately limiting the fidelity of quantum gates, quantum simulations, and other Rydberg-based quantum technologies.

Stray electric fields originating from nearby surfaces constitute one of the major sources of environmental noise for Rydberg atoms~\cite{Saffman2016Quantum, Keil10102016, PhysRevResearch.4.013207, PhysRevLett.132.113601}, and similarly limit the performance of trapped ions~\cite{PhysRevA.61.063418, RevModPhys.87.1419, 10.1063/1.5088164} and solid-state color centers~\cite{zuber2023shallow, PhysRevLett.115.087602, PhysRevB.93.024305, yuan2026surface}. Previous studies have established that adsorbates and charge accumulation on the inner surfaces of vacuum cells constitute the dominant sources of such stray electric fields~\cite{PhysRevA.69.062905, PhysRevA.75.062903, PhysRevA.81.063411, PhysRevA.84.023408, PhysRevA.86.022511, PhysRevLett.112.026101, PhysRevB.89.245435, PhysRevA.105.013107}. These surface charges generate inhomogeneous electric fields that induce Stark shifts, broaden Rydberg spectra, and degrade the coherence of Rydberg excitation.
To suppress these effects, two principal approaches have been developed. The first employs ultraviolet (UV) illumination to remove adsorbates and surface charges through light-induced desorption, thereby reducing the sources of the stray electric fields~\cite{2011Reduction, PhysRevApplied.22.064021}. The second utilizes in-vacuum electrodes to generate a compensating electric field that cancels the ambient stray field at the position of the atoms~\cite{10.1063/5.0239165}. Although these techniques have proven highly effective in mitigating stray fields associated with surface adsorbates, residual electric fields are frequently observed in experiments, indicating that additional sources of electric-field noise remain insufficiently understood.

In this work, we identify a previously unrecognized source of residual electric fields in antireflection (AR)-coated glass vacuum cells and develop an effective strategy to eliminate it. Surprisingly, we find that the conventional UV-induced charge-desorption method can increase, rather than reduce, the residual electric field in our system, providing evidence for the presence of trapped charges within the dielectric AR coating layers. To neutralize these trapped charges, we introduce a photo-ionization-based compensation scheme~\cite{zuber2023shallow, deng2026efficient}, in which additional charges are generated from a cold atomic ensemble by illuminating it with photo-ionization beams. These charges subsequently adsorb onto the vacuum-cell surface, compensating the trapped charges and thereby suppressing the stray electric field.
To verify the effectiveness of this compensation mechanism, we characterize the Stark-shift parabola of the Rydberg transition by applying controlled electric fields with external bias electrodes. After eliminating the residual stray electric field, we observe a substantial reduction in both the linewidth and frequency drift of the Rydberg transition in a single-atom array, together with an extension of the ground–Rydberg Rabi coherence time from $4.0~\mu\mathrm{s}$ to $15.8~\mu\mathrm{s}$. These results identify trapped charges in AR coating layers as a previously overlooked source of electric-field noise and establish photo-ionization-assisted charge compensation as a practical and broadly applicable approach for improving electric-field stability in neutral-atom quantum platforms.



\section{residual charges in vacuum} 
\label{2}

\begin{figure}[ptb]

\begin{center}
\includegraphics[width=8.8cm]{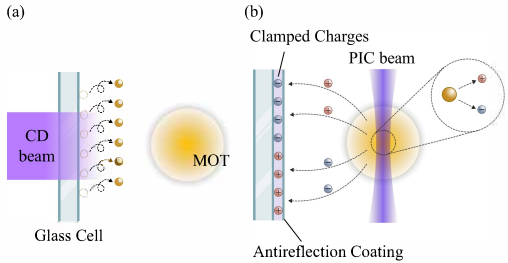}
\caption{\label{fig1}
Schematic illustration of the two strategies for mitigating stray electric fields.
(a) CD strategy. Cold atoms confined in a MOT (yellow cloud) experience stray electric fields generated by surface charges (yellow circles) on the inner wall of the vacuum cell. CD beam illumination removes these surface charges through charge desorption, thereby reducing the stray electric field.
(b) PIC strategy. In an AR-coated vacuum cell, additional charges trapped within the AR coating cannot be removed by the CD process. A PIC beam illuminates the cold atoms, generating ions and electrons that drift toward the vacuum-cell surfaces under the residual electric field and compensate the trapped charges, thereby suppressing the stray electric field.
}

\end{center}
\end{figure}

\begin{figure}[ptb]

\begin{center}
\includegraphics[width=1.0\linewidth]{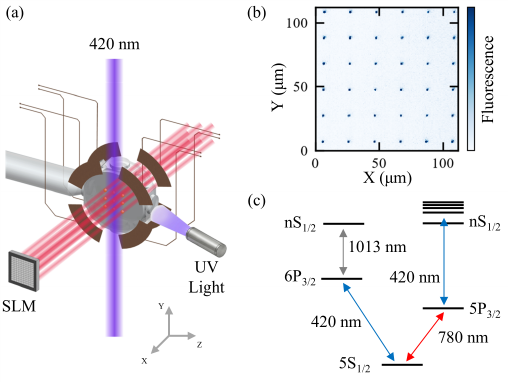}
\caption{\label{fig2}
Experimental setup and atomic energy levels.
(a) Schematic of the experimental apparatus. An 819-nm optical tweezer array is generated by a SLM and propagates along the $x$-axis to trap individual $^{87}\mathrm{Rb}$ atoms inside an octagonal fused-silica vacuum cell with AR coatings. Eight external electrodes generate controllable bias electric fields for Stark-shift measurements. A 365-nm UV beam is used for the CD process, while a 420-nm beam propagating along the $y$-axis is used for PIC.
(b) Averaged fluorescence image of a $6 \times 6$ array of individually trapped $(^{87}\mathrm{Rb})$ atoms. The nearest-neighbor spacing is $21~\mu\mathrm{m}$, sufficiently large to avoid Rydberg blockade interaction between adjacent atoms.
(c) Energy-level diagram of $^{87}\mathrm{Rb}$. Coherent excitation to the Rydberg state is achieved using 420-nm and 1013-nm laser beams with a {blue} detuning of 770 MHz from the $5S_{1/2}$ to the intermediate $6P_{3/2}$ state. During the PIC process, the same 420-nm beam, together with the 780-nm MOT beams, excites the atoms into the ionization continuum.
}
\end{center}
\end{figure}

The two types of residual charges considered in this work are illustrated in Fig.~\ref{fig1}. In an uncoated quartz vacuum cell, the dominant source of stray electric fields originates from adsorbates and charge accumulation on the inner glass surface, as illustrated in Fig.~\ref{fig1}(a). These surface-bound charges generate background electric fields that induce Stark shifts of Rydberg states and degrade the coherence of Rydberg excitation. To suppress this source of electric-field noise, the conventional charge-desorption (CD) technique employs UV illumination to remove adsorbates and surface charges from the vacuum-cell surface~\cite{PhysRevApplied.22.064021}.
For a vacuum cell coated with antireflection (AR) layers, however, an additional source of stray electric fields can arise from charges trapped within the dielectric coating, which cannot be removed by the CD process. To compensate these trapped charges, we introduce a photo-ionization compensation (PIC) strategy, as illustrated in Fig.~\ref{fig1}(b). In this scheme, a photo-ionization beam illuminates the cold atomic ensemble confined in the magneto-optical trap (MOT), generating ions and electrons through photo-ionization. Driven by the residual electric field, these charged particles migrate toward the vacuum-cell surfaces, where they compensate the trapped charges and thereby suppress the stray electric field.

The experimental setup is illustrated in Fig.~\ref{fig2}. The science chamber consists of an octagonal fused-silica vacuum cell whose inner and outer surfaces are coated with antireflection (AR) layers, as shown in Fig.~\ref{fig2}(a). A spatial light modulator (SLM) is used to generate an optical tweezer array for trapping individual $^{87}\mathrm{Rb}$ atoms. In the experiments reported here, we employ a two-dimensional $6 \times 6$ atom array with a nearest-neighbor spacing of $21~\mu\mathrm{m}$, corresponding to an overall array size of approximately $105 \times 105~\mu\mathrm{m}^2$, as shown in Fig.~\ref{fig2}(b).
To investigate the two compensation schemes, a 365-nm UV beam is used for the charge-desorption (CD) measurements, while a 420-nm laser is employed for the photo-ionization compensation (PIC) experiments. As shown in Fig.~\ref{fig2}(c), the 420-nm laser is primarily used to coherently excite atoms to the Rydberg state via a two-photon transition together with a 1013-nm laser. During the PIC process, however, the same 420-nm beam photo-ionizes trapped atoms into the continuum in combination with the 780-nm optical trapping light, thereby generating ions and electrons.


\section{Compensation of Residual charges }
\label{3}


\begin{figure}[ptb]

\begin{center}
\includegraphics[width=1.0\linewidth]{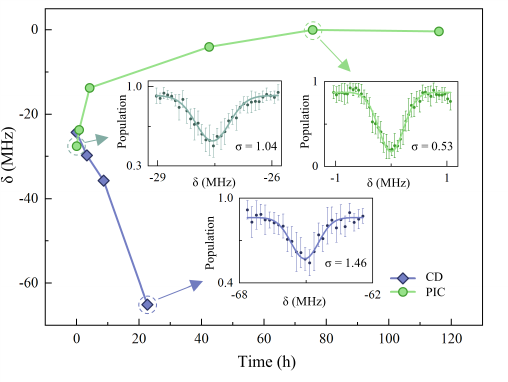}
\caption{\label{fig3}
Temporal evolution of the Rydberg transition frequency under CD and PIC condition. Blue squares correspond to continuous UV illumination, while green circles correspond to the PIC strategy. Insets show representative ground-state population spectra measured at the indicated times. The solid curves are Gaussian fits, and $\sigma$ denotes the fitted spectral linewidth of each spectrum.
}
\end{center}
\end{figure}


In this section, we evaluate the performance of the conventional CD strategy and the proposed PIC strategy using a $6 \times 6$ array of individually trapped $^{87}\mathrm{Rb}$ atoms. The atoms are initially prepared in the $\ket{5S_{1/2}, F=2, m_F=-2}$ ground state and coherently excited to the $\ket{53S_{1/2}, m_j=-1/2}$ Rydberg state through a global two-photon transition driven by 420-nm (TA-SHG-Pro, Toptica) and 1013-nm (FA-SF-1013-10-CW, Precilaser) laser beams. The initial Rydberg excitation spectrum of the atom array exhibits an overall linewidth of 1.04 MHz. This linewidth contains two contributions: the intrinsic single-atom linewidth, which is primarily limited by the excitation laser linewidth and the finite lifetime of the Rydberg state, and an additional broadening arising from spatial variations of the Rydberg transition frequency across the array due to inhomogeneous stray electric fields. Details of the experimental setup are provided in the Appendix~\ref{1}.

Figure~\ref{fig3} summarizes the temporal evolution of the ground–Rydberg transition frequency and the corresponding spectral linewidth under continuous the CD and the PIC protocol. We first investigate the conventional CD strategy by continuously illuminating the vacuum cell with a 365-nm UV LED (LBLED-0365, LBTEK). Contrary to the expected suppression of stray electric fields, the Rydberg transition frequency (blue diamonds) decreases monotonically throughout the measurement. After 20 hours of continuous UV illumination, the transition frequency has shifted by approximately 40 MHz without reaching a steady state, while the spectral linewidth increases from 1.04 MHz to 1.46 MHz. Rather than improving the electric-field environment, UV illumination therefore leads to a progressive degradation of the Rydberg spectrum. This unexpected behavior indicates that the dominant source of the residual electric field in our AR-coated vacuum cell cannot be explained solely by removable surface adsorbates. Instead, it suggests the presence of additional fixed or trapped charges that are not eliminated by the charge-desorption process, motivating the development of a compensation strategy based on generating, rather than removing, charges.

In stark contrast, under prolonged application of the PIC strategy, the Rydberg transition frequency gradually increases and eventually reaches a stable plateau after approximately 75 hours of continuous operation (green circles). Once this steady state is established, no appreciable frequency drift is observed, indicating that the residual electric field has been effectively compensated through the photo-ionization process. Concurrently, the spectral linewidth decreases by nearly 50\%, from 1.04 MHz to 0.53 MHz. This pronounced linewidth narrowing demonstrates that the PIC strategy not only stabilizes the absolute Rydberg transition frequency, but also substantially improves the spatial homogeneity of the electric field across the atom array.

The compensation is not permanent. When the PIC process is stopped after the transition frequency has stabilized, the compensated charges are gradually removed by the ion pump, allowing the trapped charges within the AR coating to progressively dominate the local electric-field environment again. Consequently, the Rydberg transition frequency slowly returns to its initial value over a timescale of several hours (see Appendix~\ref{9} for detailed measurements). To maintain a stable electric-field environment during experiments, a PIC sequence is performed prior to each measurement cycle. Specifically, the 780-nm MOT beams and the 420-nm excitation beam are simultaneously applied for 700 ms before loading atoms into the optical tweezers, thereby generating a controlled flux of photo-ionized charges that replenishes those continuously removed by the ion pump. The complete timing sequence is provided in the Appendix~\ref{8}.

For long periods without experiments, such as overnight, we additionally illuminate the MOT region with a low-power 420-nm LED (LBLED-0420, LBTEK) to provide continuous weak photo-ionization. This procedure maintains the compensated electric field environment, suppresses the gradual recovery of the stray electric field (see Appendix~\ref{9}), and reduces the operating duty cycle of the 420-nm laser, thereby extending its service lifetime. As a result, a stable electric-field environment can be reliably maintained for subsequent Rydberg experiments.


\section{Eliminated stray electric fields}
\label{4}


\begin{figure}[ptb]

\begin{center}
\includegraphics[width=1.0\linewidth]{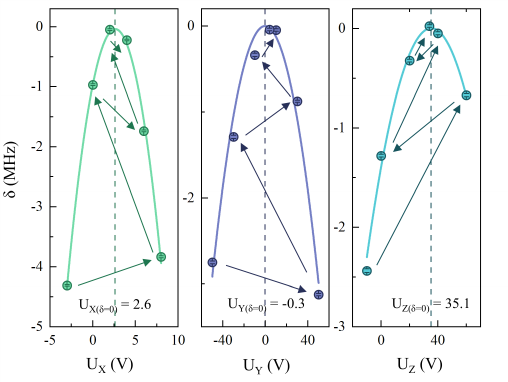}
\caption{\label{fig4}
Stark-shift characterization of the residual electric field after photo-ionization compensation. Measured Rydberg transition frequency as a function of the voltage applied along the (a) $X$-, (b) $Y$-, and (c) $Z$-directions. The definitions of the coordinate axes are shown in Fig.~\ref{fig2}(a). The experimental data (symbols) are fitted with quadratic functions (solid curves). The vertex of each parabola marked with dashed lines corresponds to the electrode voltage at which the applied electric field compensates the residual stray field along the corresponding direction.
}
\end{center}
\end{figure}

To verify that the PIC strategy effectively eliminates the residual stray electric field, we characterize the Stark shift of the Rydberg transition by applying controlled static electric fields with external electrodes. In the presence of an applied electric field ($E_A$) and a residual stray electric field ($E_S$), the Stark shift of the Rydberg energy level is given by
\begin{equation}
\delta E = -\frac{1}{2}\alpha (E_A + E_S)^2,
\end{equation}
where $\alpha$ is the polarizability of the Rydberg state. Since the applied electric field varies linearly with the electrode voltage, the Rydberg transition frequency exhibits a quadratic dependence on the applied voltage. The vertex of the resulting parabola corresponds to the condition $E_A=-E_S$, where the applied field exactly compensates the stray field at the position of the atoms. At this operating point, the transition frequency is first-order insensitive to electric-field fluctuations, providing the optimal condition for coherent Rydberg excitation.

Fig.~\ref{fig4} shows the measured Rydberg transition frequency as a function of the applied voltage along each of the three orthogonal directions. The experimental data (scatters) are well described by quadratic fits (solid lines). The vertex of each parabola determines the compensation voltage required to cancel the residual stray field along the corresponding axis. Notably, the compensation voltage required along the $Z$-axis is substantially larger than those along the $X$- and $Y$-axes. We attribute this asymmetry to the geometry of the vacuum chamber that one side of the atomic ensemble is affected by the vacuum window, whereas the opposite side is connected to the differential pumping tube, providing a smaller effective surface area for charge accumulation. Consequently, the photo-ionization-generated charges are expected to compensate the trapped charges less efficiently along this direction, leading to a larger residual field. Nevertheless, the available voltage range of the electrode system is sufficient to compensate this remaining field.

To achieve full three-dimensional compensation, we employ the iterative “scan-and-update” procedure~\cite{10.1063/5.0239165}. During each iteration, the three spatial directions are scanned sequentially. For each axis, the measured data are fitted with a parabola, the vertex voltage is extracted, and the corresponding electrode voltage is updated before proceeding to the next direction. Because the electric fields generated by the electrodes are not perfectly orthogonal, adjusting one axis slightly perturbs the compensation along the others. Consequently, several iterations are typically required before the compensation voltages converge. The procedure is terminated once the vertex positions remain unchanged within the experimental uncertainty between successive iterations.
Before applying the PIC strategy, the vertex of the Stark parabola could not even be observed. The residual stray electric field exceeded the compensation range of the available power supply, such that only one side of the parabola was experimentally accessible (see Appendix~\ref{9}). Following photo-ionization compensation, the residual electric field was sufficiently reduced to bring the vertex within the accessible voltage range, enabling measurement of the complete Stark parabola and the subsequent iterative three-dimensional compensation described above.


\section{Elongated Rydberg Rabi lifetime}
\label{5}


\begin{figure}[ptb]

\begin{center}
\includegraphics[width=1.0\linewidth]{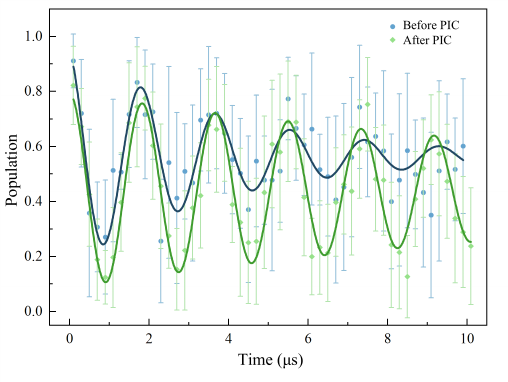}
\caption{\label{fig5}
Ground-to-Rydberg Rabi oscillations measured before and after PIC. Blue circles (green squares) represent the experimental data acquired before (after) applying the PIC strategy, while the corresponding solid curves are fits to exponentially damped sinusoidal functions. 
}
\end{center}
\end{figure}

To evaluate the improvement in coherent Rydberg control enabled by the PIC strategy, we measure ground-to-Rydberg Rabi oscillations before and after PIC, as shown in Fig.~\ref{fig5}. Before applying PIC, the Rabi oscillations are strongly damped, yielding an extracted coherence time of $4.0~\mu\mathrm{s}$. This rapid decay is primarily attributed to residual stray electric fields, which induce both temporal fluctuations of the Rydberg transition frequency and spatial inhomogeneity of the Stark shift across the atom array. After applying the PIC strategy, the oscillations become markedly more coherent, and the extracted coherence time increases to $15.8~\mu\mathrm{s}$, corresponding to an approximately fourfold improvement.

The substantial enhancement of the Rabi coherence is consistent with the linewidth narrowing reported in Sec.~\ref{3}. Prior to PIC, spatially inhomogeneous stray electric fields produce a position-dependent Stark shift across the array, causing different atoms to experience different detunings from the global excitation lasers. As a consequence, the atoms undergo Rabi oscillations with different effective frequencies and oscillation amplitudes. Averaging over the entire array therefore leads to accelerated dephasing and a reduced ensemble coherence time. Following photo-ionization compensation, the spatial variation of the Rydberg transition frequency is significantly suppressed, resulting in a much more uniform excitation condition across the array. Consequently, the ensemble Rabi oscillations exhibit substantially improved coherence. In addition to reducing spatial inhomogeneity, the PIC strategy also suppresses the long-term drift of the Rydberg transition frequency, thereby mitigating temporal dephasing during repeated experimental cycles. The combined suppression of spatial and temporal electric-field noise provides a stable environment for coherent Rydberg excitation in AR-coated vacuum cells, representing a key step toward high-fidelity quantum operations in neutral-atom quantum processors.


\section{conclusion and discussion}
\label{6}

In this work, we identify a previously overlooked source of residual stray electric fields in antireflection (AR)-coated vacuum cells and develop a photo-ionization compensation (PIC) strategy to mitigate its impact. By photo-ionizing rubidium atoms with 780-nm and 420-nm laser beams, we generate charged particles that compensate the residual electric field associated with trapped charges at the AR-coating–glass interface. As a result, the electric-field environment is significantly stabilized, leading to a reduction of the Rydberg excitation linewidth across the atom array from 1.04 MHz to 0.53 MHz. Combining PIC with electrode-based electric-field compensation further enables complete Stark-shift characterization and iterative optimization of the operating point. Consequently, the ground-to-Rydberg Rabi coherence time is extended by a factor of four, from $3.95~\mu\mathrm{s}$ to $15.84~\mu\mathrm{s}$, demonstrating a substantial improvement in coherent quantum control.

The microscopic origin of the trapped charges at the AR-coating–glass interface merits further investigation. Previous studies have shown that laser irradiation of dielectric thin films can induce long-lived negative charge accumulation~\cite{Becker:87}. In our system, repeated laser illumination may similarly lead to charge trapping within or near the AR-coating–glass interface. The dielectric multilayer structure and the associated material interfaces provide favorable conditions for charge trapping, offering a plausible explanation for both the failure of conventional UV charge desorption and the effectiveness of the PIC strategy observed in our experiments.

In summary, we demonstrate that photo-ionization compensation, combined with electrode-based electric-field cancellation, provides a practical and robust approach for stabilizing the electric-field environment in AR-coated vacuum cells. Beyond enabling long-coherence Rydberg excitation in single-atom arrays, this work identifies a previously unrecognized source of electric-field noise in dielectric-coated quantum devices and provides an effective strategy for mitigating it. We anticipate that this approach will facilitate higher-fidelity quantum operations in neutral-atom quantum processors and will be broadly applicable to other quantum platforms whose performance is limited by surface-induced electric-field noise.


\begin{acknowledgments}
We sincerely acknowledge the technical support provided by X.-D. He, P. Xu, K.-P. Wang, and Y.-B. Wang. 
C. L. thanks M.-H. Li for the valuable suggestion on photoluminescence excitation.
We acknowledge the enlightening talk given by J. Rui and instructive discussions with C. Chen, T. Chen, and S. Zhang at atomic array seminar at Wuxi in April. 
This work is supported by the National Natural Science Foundation of China (Grant Nos. 12404407, 12322408), the Guangdong Provincial Quantum Science Strategic Initiative (Grant Nos. GDZX2504005, GDZX2404001, GDZX2303006 GDZX2304002), and the Guangdong Basic and Applied Basic Research Foundation (Grant Nos. 2026A1515030019, 2024A1515012516).


\end{acknowledgments}


\appendix
\renewcommand{\thefigure}{S\arabic{figure}}  
\setcounter{figure}{0}
\section{EXPERIMENTAL SETUP}
\label{7}

\subsection{Vacuum System}

Our experiment is performed in a dual-chamber ultrahigh vacuum system designed to achieve low background gas pressure for long trapping lifetimes. The system consists of two chambers connected by a differential pumping tube.

The first chamber houses a two-dimensional magneto-optical trap (MOT) and is connected to rubidium dispensers that serve as the atomic source. This chamber is pumped by a 40 L/s ion pump and the background pressure is maintained at approximately \(10^{-7}\) Pa during operation. The 2D MOT collects and cools rubidium atoms from the dispenser flux and delivers them toward the second chamber through the differential pumping stage.

The second chamber is an octagonal fused-silica cell with both inner and outer surfaces coated with an antireflection (AR)-coating. This double-sided AR-coating is essential for efficient optical access in our experiment, as it minimizes reflection losses and maintains beam quality for the multiple laser beams used in our experiment. Critically, as discussed in the main text, this coating also introduces the buried coating-glass interface where charges can become trapped, a condition that underlies the central challenge addressed in this work. This chamber is where the 3D MOT and subsequent single atom trapping and Rydberg excitation experiments are performed. It is equipped with a titanium sublimation pump and a 75 L/s ion pump, which together maintain the background pressure at approximately \(10^{-9}\) Pa. At this pressure, the background gas collision rate is estimated to be approximately 200 s\(^{-1}\), which significantly extends the lifetime of atoms trapped in optical dipole tweezers (ODTs).

A differential pumping tube between the two chambers maintains the pressure gradient, ensuring that the high vapor pressure of the rubidium dispensers in the first chamber does not compromise the ultrahigh vacuum in the second chamber. This differential pumping stage maintains the pressure differential between the two chambers, simultaneously providing the elevated background pressure necessary for efficient 2D MOT operation and the ultrahigh vacuum environment required for the 3D MOT chamber.


\begin{figure*}[ptb]

\begin{center}
\includegraphics[width=17.8cm]{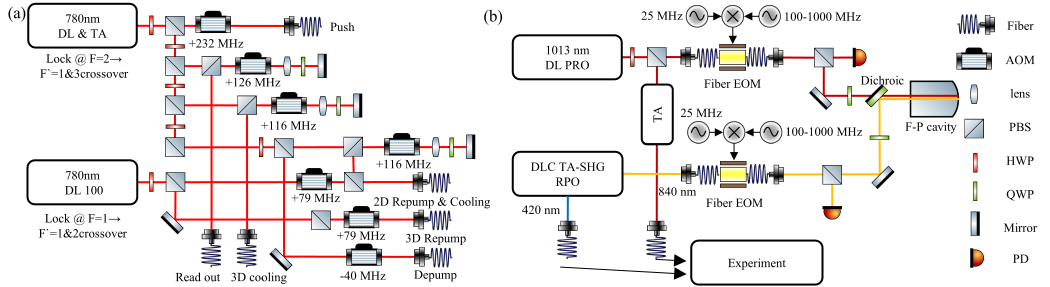}
\caption{\label{figs1}
(a) Optical layout for the 780-nm laser system, including beam splitting and delivery paths for the 2D and 3D MOT cooling beams, push beam, probe beam, optical pump beam, and read out beam. (b) PDH locking optical layout for the 420-nm and 1013-nm laser systems.}
\end{center}
\end{figure*}


\begin{figure*}[ptb]

\begin{center}
\includegraphics[width=17.8cm]{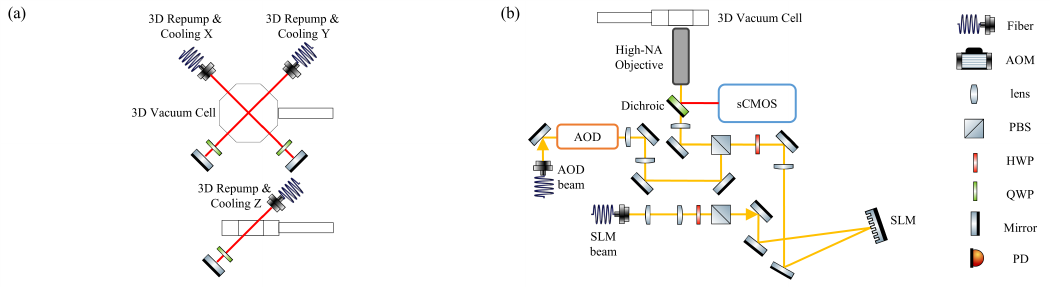}
\caption{\label{figs2}
(a) Optical layout for the 3D MOT beams entering the vacuum cell. (b) Optical layout for the 819-nm laser system, showing the beam paths for generating the optical tweezer (ODT) array via the spatial light modulator (SLM) and the movable tweezer via the acousto-optic deflector (AOD).}
\end{center}
\end{figure*}


\begin{figure}[ptb]
\begin{center}
\includegraphics[width=8.8cm]{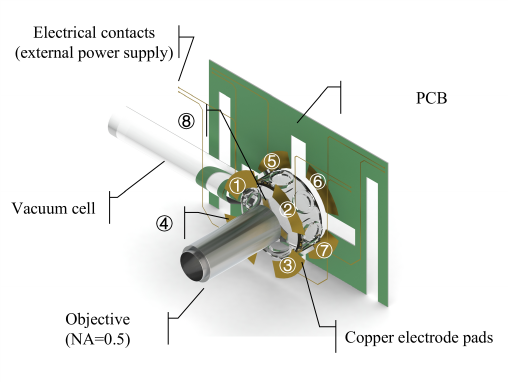}
\caption{\label{electrode}
Mechanical design and physical implementation of the electrode assembly.}
\end{center}
\end{figure}


\begin{figure}[ptb]

\begin{center}
\includegraphics[width=8.8cm]{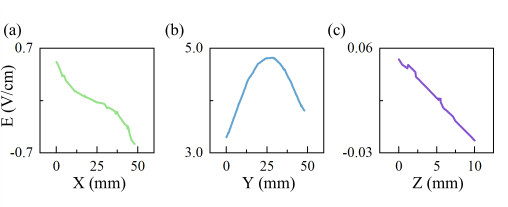}
\caption{\label{figs3}
Electrode system for stray-field cancellation. (a)--(c) Simulations of the electric-field intensity along the (a) X-axis, (b) Y-axis, and (c) Z-axis in the vicinity of the atomic array.}
\end{center}
\end{figure}


\subsection{Laser Systems}

The laser systems are a critical component of our cold-atom experiment. Not only do they provide the cooling light for MOTs, but also play a central role in multiple stages of our experiment, including atom transport, state preparation, energy level manipulation, and quantum state read out.

In order to generate $^{87}$Rb MOT and realize the state preparation, we employ two 780-nm laser systems, which drive the \(5S_{1/2} \leftrightarrow 5P_{3/2}\) transition of \(^{87}\)Rb. The first is a tapered amplifier (TA) laser (HPL-780F, UniQuanta), whose frequency is stabilized using saturated absorption spectroscopy. This laser generates multiple beams for different purposes, providing the cooling light (detuned by -12 MHz from the \(F=2 \leftrightarrow F'=3\) transition) for both the 2D and 3D MOTs, the optical pump beam for state preparation, the read out beam (resonant with the \(F=2 \leftrightarrow F'=3\) transition) for quantum state detection, the probe beam for atomic fluorescence imaging, and the push beam (detuned by -12 MHz from the \(F=2 \leftrightarrow F'=3\) transition) that transfers atoms from the 2D MOT chamber to the 3D MOT chamber. The second laser is a digital laser (DL, Toptica), which is also locked via saturated absorption spectroscopy and provides the repump light (resonant with the \(F=1 \leftrightarrow F'=2\) transition) for both MOTs. The detailed beam splitting and delivery paths to the vacuum cell are shown in Fig.~\ref{figs1}(a) and Fig.~\ref{figs2}(a), respectively. Typical optical powers used for MOTs are as follows: the 2D MOT cooling beam operates at 100 mW, the 3D MOT cooling beam at 5 mW, the push beam at 10 mW, the read out beam at 200 $\mu$W, the 2D MOT repump beam at 5 mW, and the 3D MOT repump beam at 10 mW. The magnetic field for the 3D MOT is provided by a pair of anti-Helmholtz coils, generating a field gradient of approximately 9.8 Gs/cm at a current of 3 A. A separate pair of coils provides the magnetic field for the 2D MOT, with a gradient of approximately 10 Gs/cm at the same current. With the combination of the cooling light, repump light, and the magnetic field gradient, the 2D and 3D MOTs are formed.

In addition to MOT generation, the 780-nm laser system is also used for quantum state preparation and state detection. For internal state manipulation, we employ probe and optical pumping beams, which are coupled into a single optical fiber containing three spectral components. The first component has the same frequency as the cooling beam, the second component matches the repump light frequency, and the third component is a depump beam resonant with the \(F=2 \leftrightarrow F'=2\) transition. The typical powers for these three components are 100 \(\mu\)W, 5 \(\mu\)W, and 2.5 \(\mu\)W, respectively. The probe beam, consisting of the first two components, continuously drives atoms from the ground state to the excited state, and the subsequent spontaneous emission produces fluorescence that is collected by a sCMOS camera (Dhyana 400BSI V3, TUCSEN) for imaging. Compared with direct MOT light imaging, the probe beam introduces lower noise and less heating, making it preferable for repeated fluorescence imaging throughout the experimental sequence. The optical pump beam, consisting of the second and third components, prepares atoms into different initial states by manipulating the laser polarization. In this work, using \(\sigma^-\) polarized 780-nm light, we initialize atoms in the $\ket{5S_{1/2}, F=2, m_F=-2}$ state. After a series of laser manipulations, the atoms eventually return to the ground state. The read out beam, resonant with the \(F=2 \leftrightarrow F'=3\) transition, is then applied to selectively remove atoms in the \(F=2\) manifold. By measuring the remaining atom population, we determine the final internal state of the atoms.

819-nm laser system is employed for single atoms trapping and rearrangement. The laser consists of a seed laser injected into a TA. As illustrated in Fig.~\ref{figs2}(b), the output beam is shaped by a spatial light modulator (SLM) and focused through a high-numerical-aperture objective (NA = 0.5, shown in Fig.~\ref{electrode}) to generate a $6\times6$ tweezer array for trapping single atoms, with a single-atom loading probability of approximately 50\%. The corresponding fluorescence image of the $6\times6$ atom array is shown in Fig.~\ref{fig2}(b), which is averaged over 100 experimental shots to clearly resolve the occupied sites. To minimize the effects of Rydberg interaction, the spacing between adjacent atoms is set to $21\ \mu\text{m}$. In addition, an acousto-optic deflector (AOD) is used to produce a movable optical tweezer for moving atoms in the array, generating a defect-free configuration.

For Rydberg transition, we employ 420-nm and 1013-nm laser systems. The 420-nm laser (TA-SHG-Pro, Toptica) is generated by frequency doubling an amplified 840-nm laser. The 1013-nm seed laser (DL Pro, Toptica) is injected into a tapered amplifier (FA-SF-1013-10-CW, Precilaser). The maximum powers of 420-nm and 1013-nm lasers are 1 W and 13.5 W, respectively. At the position of the atoms, the 420-nm beam has a waist radius of approximately 280 \(\mu\)m, while the 1013-nm beam has a waist radius of approximately 115 \(\mu\)m. Both the 840-nm and 1013-nm lasers are frequency-stabilized using the Pound–Drever–Hall (PDH) technique. In the PDH scheme, we adopt an electro-optic dual-sideband (EDSB) method, in which two RF signals of different frequency are combined using a splitter and fed into a single fiber electro-optic modulator (EOM). One of the RF signals provides the modulation for PDH locking, while the other serves as a frequency shifting tone for laser frequency scanning. This approach allows both frequency modulation and wide scanning to be achieved with a single EOM, simplifying the optical layout. With this locking scheme, the linewidth of both lasers is $\sim$100 Hz. The PDH locking and experimental optical paths are illustrated in Fig.~\ref{figs1}(b). The theoretical maximum single-photon Rabi frequencies for the 420-nm and 1013-nm transitions are both approximately 110 MHz, and the corresponding two-photon Rabi frequency is approximately 7.8 MHz with a blue detuning of 770 MHz for the ground-to-intermediate transition.


\subsection{Electrode system}

To compensate for stray electric fields, we design an external segmented electrode system consisting of two parallel printed circuit boards (PCBs), each embedding four arc-shaped electrodes, forming a total of eight electrodes. The two PCBs are mounted on the front and rear sides of the octagonal vacuum cell, respectively, with dimensions of 200 mm \(\times\) 139 mm each. The distance between two PCBs is 37 mm. The geometric center axis of the electrode assembly is aligned with the center of the vacuum cell, ensuring front-rear symmetry with respect to the atomic position. Each PCB contains four independently controlled arc-shaped electrodes, forming a total of eight voltage channels. The arc-shaped electrode geometry is chosen to match the circular windows of the vacuum cell: each segment has an angular span of 60\(^\circ\) and a radial width of 15 mm. This design allows the electrodes to be positioned close to the atomic region while maintaining optical access through the windows, and also reduces local field concentration that would otherwise arise from sharp corners. Thus, the structure achieves a balance between effective electrode area and optical transparency.

Figure~\ref{figs3}(d) shows the mechanical design and physical implementation of the electrode assembly, including the electrode positioning and mounting relative to the vacuum cell. The electrode arrangement allows independent voltage control along each axis, enabling the iterative "scan-and-update" procedure for locating the vertex of the Stark-shift parabola.

The eight electrodes are driven by independent bipolar dc voltage channels with a nominal operating range of \(\pm 80\) V. The voltages applied to the eight electrodes, denoted as \(U_1\) - \(U_8\) (with the numbering defined in Fig.~\ref{figs3}(d)), are related to the desired basis voltages \(U_X\), \(U_Y\), and \(U_Z\) by the following linear combinations:

\begin{align}
U_1 &= \frac{1}{2}(U_X + U_Y - U_Z), \\
U_2 &= \frac{1}{2}(U_X + U_Y + U_Z), \\
U_3 &= \frac{1}{2}(U_X - U_Y + U_Z), \\
U_4 &= \frac{1}{2}(U_X - U_Y - U_Z), \\
U_5 &= \frac{1}{2}(-U_X + U_Y - U_Z), \\
U_6 &= \frac{1}{2}(-U_X + U_Y + U_Z), \\
U_7 &= \frac{1}{2}(-U_X - U_Y + U_Z), \\
U_8 &= \frac{1}{2}(-U_X - U_Y - U_Z).
\end{align}

This approach enables precise three-dimensional control of the electric field, with the electrode voltage distribution tailored to compensate for the stray field at the atomic position. The simulated electric field components at the atomic position are related to the basis voltages by the following calibration matrix:

\begin{align}
E_X &= -22.46\,U_X + 0.38\,U_Y - 2.97\,U_Z, \\
E_Y &= -0.22\,U_X - 24.21\,U_Y + 0.15\,U_Z, \\
E_Z &= 0.199\,U_X - 0.11\,U_Y - 25.25\,U_Z,
\end{align}
where \(E_X\), \(E_Y\), and \(E_Z\) are in units of V/m. The diagonal coefficients represent the direct coupling between each basis voltage and its corresponding field component, while the off-diagonal terms account for cross-coupling due to the geometric asymmetry of the electrode structure. Any arbitrary static electric field direction can be achieved through linear superposition of these basis fields, enabling full three-dimensional stray-field compensation.

The "scan-and-update" procedure is implemented as follows. First, we scan the X direction by varying \(U_X\) while setting \(U_Y = U_Z = 0\). The voltage \(U_X\) that minimizes the Rydberg transition frequency is extracted from a parabolic fit and fixed at that value. Next, we scan the Y direction by varying \(U_Y\) while keeping \(U_X\) fixed at its optimized value and \(U_Z = 0\). The vertex voltage for \(U_Y\) is then extracted and fixed, with both \(U_X\) and \(U_Y\) now held at their optimized values. Finally, we scan the Z direction by varying \(U_Z\) while keeping both \(U_X\) and \(U_Y\) fixed at their optimal values determined previously. The vertex voltage for \(U_Z\) is extracted and fixed, completing a full iteration. Since the three axes are not perfectly orthogonal in terms of the electrode response, several iterations may be required to converge to the global minimum. This ensures that, during the scan of a subsequent direction, the compensation along previously optimized axes remains active. 

The design and simulated performance of the electrode system are shown in Fig.~\ref{figs3}. Figures~\ref{figs3}(a)--\ref{figs3}(c) present the simulations of the electric-field intensity distribution along the X-, Y-, and Z-axes, respectively, in the vicinity of the atomic array. The simulations confirm that the electrode configuration produces a sufficiently uniform electric field at the position of the atoms, enabling precise control of the stray-field compensation. The spatial variation of the field across the array region is minimal, ensuring that all atoms in the array experience a nearly identical applied field.

It is worth noting that the actual electric field experienced by the atoms is considerably smaller than the simulated values. A plausible explanation for this discrepancy is that the applied electric field drives the adsorbed rubidium atoms or ions on the cell walls to migrate and redistribute, thereby generating an internal field that partially cancels the applied field. Quantitatively, the measured field along the X-direction is approximately 43.62\% of the simulated value, whereas the measured fields along the Y- and Z-directions are only 3.78\% and 3.55\% of their simulated values, respectively. The significantly weaker screening along the X-direction may arise from the geometrical difference in the glass surfaces perpendicular to each axis. The octagonal cell has two large parallel windows facing the X-direction, providing a large surface area for adsorbate redistribution. When the field drives adsorbate motion, the induced charge redistribution occurs over a large area, resulting in a relatively small local screening field. In contrast, the glass surfaces perpendicular to the Y- and Z-directions are considerably smaller, and the same field-driven adsorbate migration is confined to a limited area, which can produce a stronger local cancellation field. This geometric factor could explain the order-of-magnitude difference in screening between the X- and the Y/Z-directions. Despite this reduction, the electrode system still provides sufficient control to perform the iterative cancellation procedure described in the main text.


\section{EXPERIMENTAL SEQUENCE AND DATA PROCESS}
\label{8}


\begin{figure}[ptb]

\begin{center}
\includegraphics[width=8.8cm]{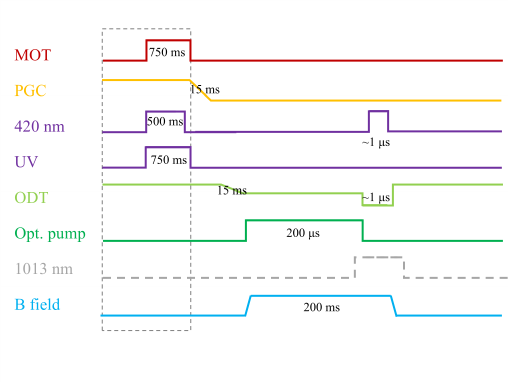}
\caption{\label{figs4}
Experimental sequence diagram. The CD and PIC stages are enclosed by dashed boxes.}
\end{center}
\end{figure}


The performance of charge-desorption (CD) and photo-ionization compensation (PIC) schemes is characterized through Rydberg excitation experiments performed in a $6\times6$ randomly loaded square array of single atoms as illustrated in Fig.~\ref{fig2}(b).

As discussed in the main text, when PIC is turned off, the Rydberg excitation frequency gradually drifts back toward its initial value as the ion pump gradually removes the ions responsible for charge compensation. Therefore, to maintain the stabilized electric-field environment during data acquisition, the PIC sequence must be incorporated into the experimental timing. The experimental sequence is illustrated in Fig.~\ref{figs4}, with the duration of each stage indicated in the figure. The UV illumination and ionization stages are enclosed by dashed boxes, indicating that these steps are applied conditionally depending on the specific stabilization scheme being tested. Following these stages, the sequence proceeds with polarization gradient cooling (PGC) and adiabatic reduction of the trap depth to lower the atomic temperature, followed by state preparation and Rydberg excitation.

Rydberg atoms are repelled by the 819-nm optical tweezer potential, allowing the excitation probability to be determined by monitoring the presence or absence of atoms in the tweezers after the excitation pulse: atoms that remain in the trap are in the ground state, while those ejected are inferred to have been excited to the Rydberg state.

For Rydberg spectroscopy, the transition frequency is scanned while the atom loss probability is recorded. In our experiment, this is achieved by scanning the frequency of the 840-nm EOM while keeping the 1013-nm laser frequency fixed. Since the 840-nm light is frequency-doubled to generate the 420-nm laser, the actual frequency shift of the 420-nm light is twice that of the 840-nm EOM drive frequency. For Rabi oscillation measurements, the pulse duration is varied. At each frequency (or time) data point, the measurement is averaged over ten experimental runs. The resulting Rydberg transition spectra are fitted to Gaussian functions to extract the center frequency and linewidth, while the Rabi oscillations are fitted to exponentially damped sinusoidal functions to extract the Rabi frequency and coherence time.


\section{HOMOGENEITY OF RYDBERG TRANSITION FREQUENCY}
\label{10}

\begin{figure*}[ptb]

\begin{center}
\includegraphics[width=17.8cm]{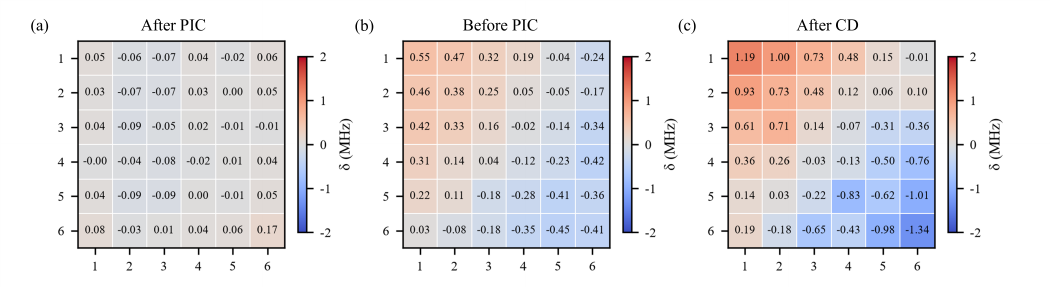}
\caption{\label{frequency_heatmap}
Spatial distribution of the Rydberg excitation frequency across the $6\times6$ atom array under different conditions: (a) after PIC, (b) before PIC, and (c) after CD. The three conditions correspond to those shown in the insets of Fig.~\ref{fig3}. Each cell represents a single atom, and the relative frequency shift $\delta = f_{\text{site}} - \bar{f}$, where $f_{\text{site}}$ is the transition frequency of a given site and $\bar{f}$ is the average frequency over all occupied sites in the array.
}
\end{center}
\end{figure*}

To further visualize the spatial homogeneity of the electric field across the array, we examine the Rydberg transition frequency distribution of individual atoms under the three conditions corresponding to Fig.~\ref{fig3}. The measurements are performed with the 420-nm and 1013-nm laser powers at the atoms of 19 mW and 640 mW, respectively, corresponding to single-photon Rabi frequencies of 37.4 MHz and 22 MHz. Figure~\ref{frequency_heatmap} presents heatmaps of the $6\times6$ array after PIC, before PIC, and after CD. Each cell represents a single atom, and the color indicates the relative frequency shift $\delta = f_{\text{site}} - \bar{f}$, where $f_{\text{site}}$ is the transition frequency of a given site and $\bar{f}$ is the average frequency over all occupied sites in the array.

Before PIC [Fig.~\ref{frequency_heatmap}(b)], a clear spatial gradient is observed: the excitation frequency gradually decreases from the upper-left corner to the lower-right corner of the array. The frequency difference between the upper-left and lower-right regions is approximately 1 MHz, which directly contributes to the broadening of the overall array linewidth. This gradient indicates that the stray electric field varies significantly across the spatial extent of the array, causing atoms at different positions to experience different Stark shifts. After CD [Fig.~\ref{frequency_heatmap}(c)], the gradient persists and the frequency difference between the upper-left and lower-right regions increases to approximately 1.3 MHz, consistent with the overall linewidth broadening observed in Fig.~\ref{fig3}, indicating that CD does not eliminate the underlying spatial inhomogeneity.

In stark contrast, with PIC [Fig.~\ref{frequency_heatmap}(a)], the frequency gradient across the array is drastically suppressed. The frequencies become tightly clustered, with the spatial variation reduced to a level comparable to the fitting uncertainty. This demonstrates that PIC not only reduces the overall linewidth but also largely eliminates the spatial inhomogeneity of the stray electric field, ensuring that all atoms in the array experience a nearly identical electric-field environment. The residual gradient is attributed to two factors. First, the finite beam waist of the excitation laser introduces a spatial variation in the Rabi frequency across the array. Second, the 420-nm and 1013-nm laser beams are not perfectly overlapped in space, which leads to position-dependent relative intensities across the array and further contributes to the spatial inhomogeneity of the effective coupling strength. {Quantitatively, the AC Stark shifts induced by the 420-nm and 1013-nm lasers at the center of the beam waist are $\delta_{420}(r=0) = 0.454$ MHz and $\delta_{1013}(r=0) = 0.157$ MHz, respectively. At a radial distance of $r=80$ $\mu$m from the center, the corresponding shifts are reduced to $\delta_{420}(r=80) = 0.328$ MHz and $\delta_{1013}(r=80) = 0.023$ MHz. These variations, together with the finite beam waist, imperfect overlap of the two beams, and the fact that the atom array is not perfectly positioned at the beam waist, result in a residual frequency gradient across the array. Nevertheless, the frequency variation across all sites remains within approximately 0.1 MHz, confirming that the spatial homogeneity of the electric field after PIC is already very good and that the residual gradient does not compromise the overall performance.}

{Note that the atom array was not rearranged into a defect-free configuration. As a result, a few sites with inherently low single-atom loading probabilities are present in the array. In addition, the number of averages per data point is limited. Consequently, these sites have larger fitting uncertainties and deviate in frequency from the majority of atoms. However, they do not affect the overall conclusion regarding the spatial homogeneity of the electric field, as the dominant trend across the array remains clear.}

The restoration of frequency homogeneity by PIC has important implications for quantum operations. In a spatially inhomogeneous field, atoms at different positions experience different detunings from the global excitation laser, leading to nonuniform Rabi frequencies and reduced ensemble-averaged coherence. By restoring spatial homogeneity, PIC ensures that all atoms respond identically to global control pulses, which is essential for scalable quantum information processing with atom arrays.


\section{LONG-TERM STABILITY}
\label{9}

\begin{figure*}[ptb]

\begin{center}
\includegraphics[width=17.8cm]{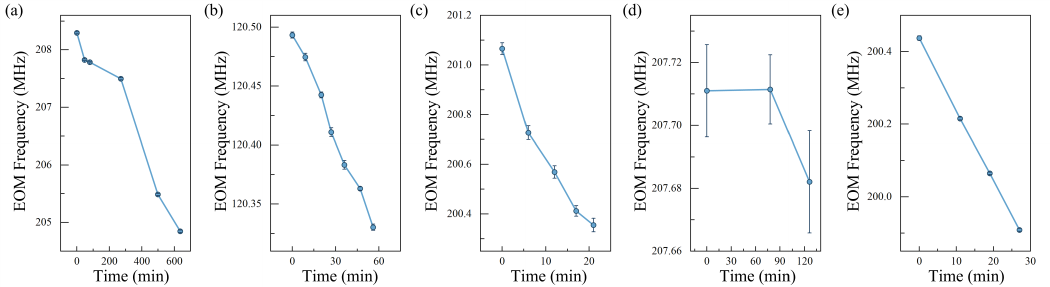}
\caption{\label{figs5}
Rydberg transition frequency as a function of time under different conditions: 
(a) PIC turned off after stabilization.
(b) $43S_{1/2}$ state, with UV illumination applied in the sequence. 
(c) $53S_{1/2}$ state, with UV illumination applied in the sequence. 
(d) $53S_{1/2}$ state, with photo-ionization applied in the sequence. 
(e) $53S_{1/2}$ state, without photo-ionization applied in the sequence.
}
\end{center}
\end{figure*}

\begin{figure*}[ptb]

\begin{center}
\includegraphics[width=17.8cm]{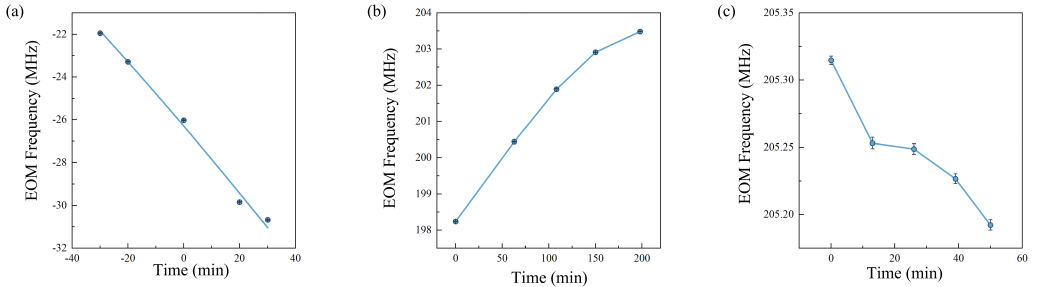}
\caption{\label{figs7}
(a) Parabolic scan of the Rydberg transition frequency as a function of the applied voltage along the Y-axis under continuous UV illumination. The dots represent the experimental data, and the solid curve is a parabolic fit to the data.
(b) Rydberg transition frequency as a function of time when the 420-nm LED is applied within the experimental sequence.
(c) Rydberg transition frequency as a function of time when the 420-nm LED is applied outside the sequence.
}
\end{center}
\end{figure*}


We first note that the compensation effect of PIC is not permanent. Once the Rydberg transition frequency is stabilized via the PIC strategy, it exhibits gradual drift toward its initial value upon cessation of ionization. As shown in Fig.~\ref{figs5}(a), the frequency recovers by approximately 3.5 MHz within 11 hours after photo-ionization is turned off. This indicates that the ions responsible for charge compensation are gradually removed by the ion pump, and the trapped interfacial charges regain their influence on the electric-field environment. To maintain the stabilized condition, continuous or periodic ionization is required.

To further demonstrate the different behaviors of CD and PIC in our AR-coated vacuum cell, we measure the time evolution of the Rydberg transition frequency during repeated experimental runs under continuous application of these stabilization schemes in the experimental sequence. The measurements are performed on different Rydberg states to compare their sensitivities to electric-field noise. The results are shown in Fig.~\ref{figs5}.

Figure~\ref{figs5}(b) presents the transition frequency of the $43S_{1/2}$ state under continuous UV illumination in the sequence. Over a period of 60 minutes, the frequency drifts by approximately 170 kHz. Figure~\ref{figs5}(c) shows the same CD protocol applied to the $53S_{1/2}$ state. In this case, the frequency drift is significantly faster: within only 20 minutes, the frequency shifts by about 700 kHz. This rapid drift is consistent with the higher polarizability of the $53S_{1/2}$ state compared with the $43S_{1/2}$ state, as the sensitivity to electric fields scales strongly with the principal quantum number $n$. Consequently, the same stray electric-field noise induces a much larger frequency shift for higher-lying Rydberg states.

In stark contrast, when PIC is applied to the $53S_{1/2}$ state, as shown in Fig.~\ref{figs5}(d), the transition frequency exhibits only a small drift of approximately 30 kHz over 120 minutes of continuous experimental runs. After about 75 minutes, the frequency stabilizes and shows negligible further variation. This demonstrates that PIC effectively suppresses the electric-field noise in our AR-coated cell, even for highly sensitive Rydberg states, whereas UV photodesorption fails to stabilize the frequency and instead exacerbates the drift. The contrast between the CD and PIC results confirms that PIC is a robust method for maintaining a stable electric-field environment in AR-coated vacuum cells.

In addition, we also investigate the stability of the compensated state after PIC has been established. Figure~\ref{figs5}(e) shows the evolution of the Rydberg transition frequency when no ionization is applied during the experimental sequence after PIC has been completed. Within 30 minutes, the frequency drifts by approximately 500 kHz, drifting back toward its initial uncompensated value. This relatively rapid drift is even more pronounced than the UV-induced drift observed in Figs.~\ref{figs5}(b) and \ref{figs5}(c), despite the different timescales, because the frequency starts from the stabilized PIC value and returns to the initial value with a steep slope. This observation confirms that the compensation effect is not permanent: once PIC is stopped, the ions responsible for charge compensation are gradually removed by the ion pump, and the trapped interfacial charges regain their influence on the electric-field environment. This further motivates the need for continuous or frequent ionization during extended idle periods, such as the 420-nm LED maintenance scheme applied between experimental sessions, or the ionization step incorporated into the experimental sequence, as described above, to sustain the compensated state over extended periods.

To further confirm the ineffectiveness of CD scheme in our AR-coated cell, we attempted to perform the electrode-based parabolic scan under continuous UV illumination. As shown in Fig.~\ref{figs7}(a), only one side of the parabola was accessible within the voltage range of our power supply, and the vertex could not be resolved. This indicates that the residual stray electric field under CD scheme is too large to be eliminated by the electrodes. This observation is consistent with the frequency drift shown in Figs.~\ref{figs5}(b) and \ref{figs5}(c), and the linewidth broadening shown in Fig.~\ref{fig3}, further supporting the conclusion that CD cannot desorb trapped interfacial charges in the AR-coated cell.


To further investigate the role of the 420-nm LED in maintaining the compensated electric-field environment, we compared the time evolution of the Rydberg excitation frequency when the 420-nm LED was used as the ionization source instead of the TA laser, both within and outside the experimental sequence. As shown in Fig.~\ref{figs7}(c), when the LED is applied only within the experimental sequence (i.e., during the short ionization window), the transition frequency continuously decreases, heading back toward its initial value before ionization, without showing any sign of stabilization. This indicates that the limited exposure time within the sequence is insufficient to generate enough ions to compensate for the trapped charges. In contrast, when the LED is left on continuously between experimental runs, as shown in Fig.~\ref{figs7}(b), the excitation frequency gradually increases and exhibits a tendency toward saturation over time. This suggests that the extended illumination period allows the LED to accumulate a sufficient number of ions to partially neutralize the trapped interfacial charges. The difference between the two cases is attributed to the relatively low power density of the focused LED: while the LED can generate ions through photo-ionization, its ionization rate is much lower than that of the TA laser. Consequently, only prolonged exposure provides sufficient integrated ion flux to maintain the compensated field environment.

\bibliography{reference}

\end{document}